# Predicting Pre-monsoon Thunderstorms -A Statistical View through Propositional Logic

Surajit Chattopadhyay**

11/3B, Doctor Lane,

Kolkata-700014

E-mail   surajit_2008@yahoo.co.in

---

---

## Abstract

Thunderstorms very close to a monsoon and not so close to a monsoon are considered in this analysis. Some important predictors are considered. Pearson Correlation Coefficient and lag-1 autocorrelation coefficients are calculated to create necessary universes of discourse for prepositional logic. The purpose is to make regression analysis more convenient for prediction of the pre-monsoon thunderstorm weather phenomenon.

** Presently:   1/19 Dover Place, Kolkata 700 019, West Bengal, India

## 1. Introduction

The severe thunderstorm is a very important weather phenomenon in Gangetic West Bengal (GWB) during March to May. Thunderstorms of this period are called pre-monsoon thunderstorms. The present study encompasses some important parameters associated with pre-monsoon thunderstorms and tries to identify one or more important predictors and predictands so that in the future, regression analysis (simple or multiple) can be done conveniently to study pre-monsoon thunderstorms. Moreover, the study further tries to understand whether a particular predictor can be used with the same predictand to analyze the thunderstorms very close to a monsoon as well as thunderstorms not very close to a monsoon. Because the dataset is vastly complex, the study has relied on the kind of robust summary offered by a linguistic proposition applied to statistical measures. Two consecutive months of pre-monsoon season are considered in this study.

## 2. Data

The dataset consists of the values of some parameters associated with severe thunderstorms of April and May occurring over GWB between 1987 and 1998. The total number of thunderstorms considered in this study is 130. Parameters considered for this study are:

- Duration (d) of the thunderstorm
- Change in air pressure (AP) during the thunderstorm
- Change in surface temperature (A T) during the thunderstorm
- Maximum wind speed (v) associated with the thunderstorm
- Change in relative humidity (A R/H) during the thunderstorm.

## 3. Methodology

The methodology adopted in the present study consists of

- Calculation of Pearson Correlation Coefficient (PCC)
- Testing for persistence
- Prepositional logic

*3.1 Pearson Correlation Coefficient (PCC)*

Pearson Correlation Coefficient (PCC) measures the degree of association between two variables 'x' and 'y'. Mathematically PCC is defined as

$$\rho_{xy} = \frac{\frac{1}{n}\sum_{i=1}^{n}(x_i - \bar{x})(y_i - \bar{y})}{\sqrt{\frac{1}{n}\sum_{i=1}^{n}(x_i - \bar{x})^2}\sqrt{\frac{1}{n}\sum_{i=1}^{n}(y_i - \bar{y})^2}} \quad (1)$$

where, $n \rightarrow$ Total number of observations

$\bar{x} \rightarrow$ Mean of the variable $x$

$\bar{y} \rightarrow$ Mean of the variable $y$

In the present paper the following

$\rho_{d\Delta p}$, $\rho_{d\Delta T}$, $\rho_{dv}$, $\rho_{d\Delta R/H}$, $\rho_{v\Delta T}$, $\rho_{v\Delta P}$, $\rho_{v\Delta R/H}$, $\rho_{\Delta R/H\Delta P}$, $\rho_{\Delta R/H\Delta T}$, and $\rho_{v\Delta R/H}$, The aforesaid quantities are calculated separately for April and May for each year.

*3.2 Testing for persistence*

Persistence means existence of statistical dependence among successive values of the same variable (Wilks, 1995). Persistence is measured by lag-1 autocorrelation .

In the present paper, PCC's mentioned in section 3.1 are considered as a dichotomous random variable X defined as

$$X = 1 \quad \text{if } |p| < 0.5$$
$$= 0 \quad \text{if } |p| > 0.5$$

Then, sequences of entries 0 and 1 are constructed separately for April and May. In the next step, the lag-1 autocorrelation coefficient is calculated for each sequence.

*3.3 Prepositional logic*

Prepositional logic is a generalized logic that includes all possible values between 0 and 1. In this logic, a relationship is required to express the distribution of the truth of a variable (Klir and Folger, 2000). A function called a "membership function" is needed to indicate the extent to which a variable 'x' has the attribute 'F'. Membership functions are defined on a universe of discourse indicated by the research variable.

In the present paper, lag-1 autocorrelations create the required universes of discourse. XI is the universe of discourse for April and X2 is the universe of discourse for May. The proposition 'P' tested for the present study is:

"The degree of association between any pair of parameters is consistently very high is very true". Thus, the membership function is framed as

$$\mu_P(x) = 0 \quad \text{for } x < 0.5$$
$$x/0.6 \quad \text{for } 0.5 < x < 0.6 \quad (2)$$
$$1 \quad \text{for } x > 0.6$$

## 3. Result and discussion

From the above study the following are found (Fig.l) to be highly true:

(a) In the month of April, maximum wind speed associated with a severe thunderstorm is mostly dependent upon the change in air pressure during the thunderstorm. But, the dependence is less in the month of May.

(b) In the month of May, maximum wind speed associated with a thunderstorm depends mostly upon the change in relative humidity during the thunderstorm. Whereas, in April, maximum wind speed has no relationship with the change in relative humidity during the thunderstorm.

(c) Change in the surface temperature during a thunderstorm depends highly upon duration of the thunderstorm in April. But, in May, they have almost no association.

(d) In the month of May, change in relative humidity during thunderstorms depends upon the duration of the thunderstorm. Whereas, in April, these two parameters have no association.

(e) Degree of association between change in air-pressure and change in relative humidity remains almost the same in the months of April and May.

## 4. Conclusion

From the above study it can be concluded, in the study area, that if maximum wind speed associated with a severe thunderstorm is considered as a predictand, then change in air pressure is a good predictor in the month of April and change in relative humidity is a good predictor in the month of May. Duration of a thunderstorm can be used as a good predictor with change in surface temperature as predictand in the month of April and with change in relative humidity as predictand in the month of May. It can further be concluded that relation between change in air pressure and change in relative humidity does not change in spite of advancement of monsoon.

## 5. Acknowledgement

The author wishes to thank Prof Sutapa Chaudhuri, Coordinator, Department of Atmospheric Sciences, University of Calcutta for providing the RMO data required for this research.

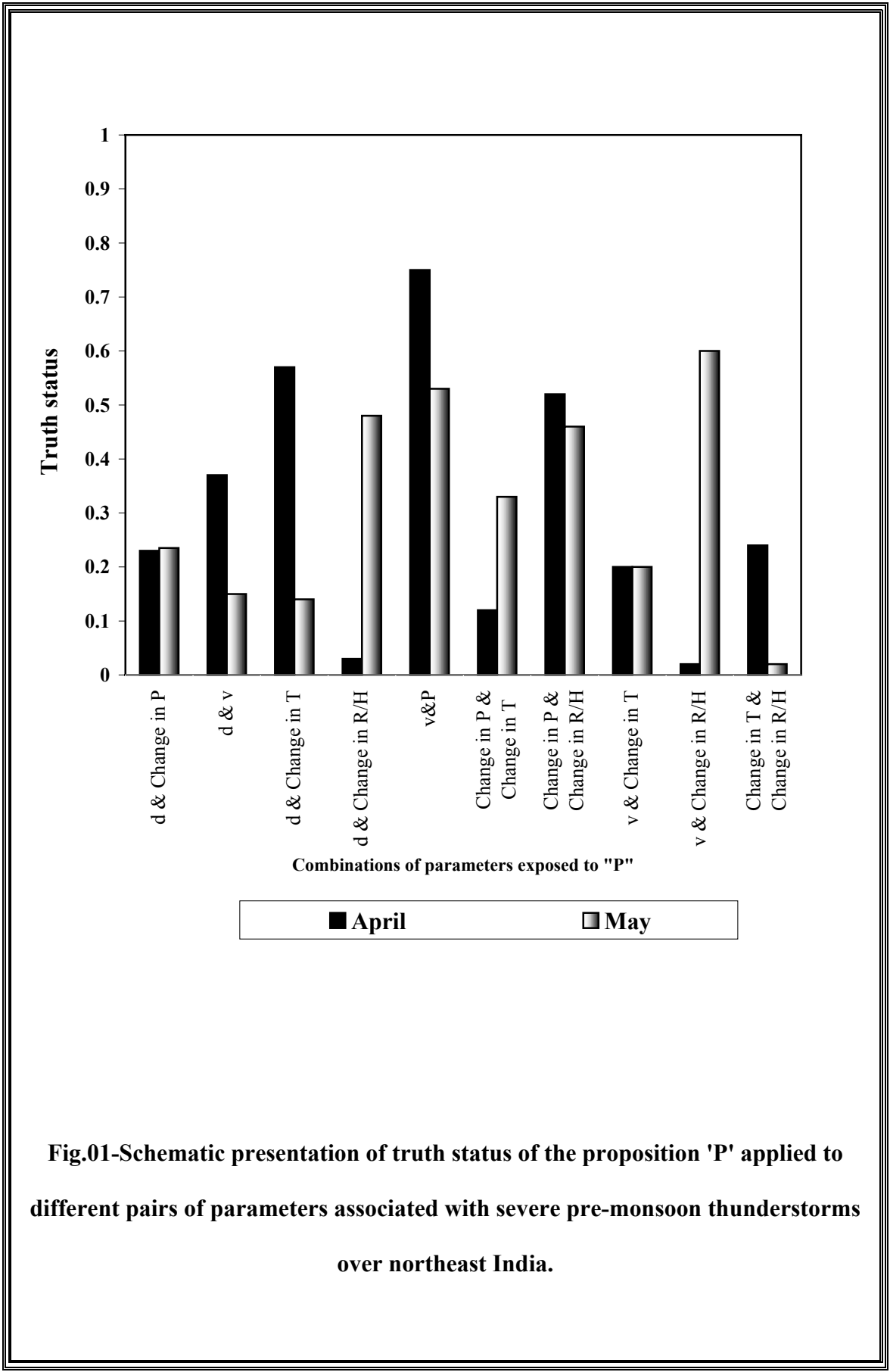

**Fig.01-Schematic presentation of truth status of the proposition 'P' applied to different pairs of parameters associated with severe pre-monsoon thunderstorms over northeast India.**